\documentclass[10pt]{iopart}
\usepackage{graphicx}
\usepackage[usenames]{color}

\begin{document}

\title[Microstability analysis of pellet fuelled discharges in MAST]
{Microstability analysis of pellet fuelled discharges in MAST}

\author{L. Garzotti$^{1}$, J. Figueiredo$^{2}$, C. M. Roach$^{1}$, M. Valovi\v{c}$^{1}$,
D. Dickinson$^{1}$, G. Naylor$^{1}$, M. Romanelli$^{1}$, R. Scannell$^{1}$, G. Szepesi$^{1,3}$ and the MAST team}

\address{$^{1}$Euratom/CCFE Fusion Association, Culham Science Centre, Abingdon, Oxon, OX14 3DB, UK}
\address{$^{2}$Association EURATOM-IST, Av. Rovisco Pais, Lisboa 1049-001, Portugal}
\address{$^{3}$Centre for Fusion, Space and Astrophysics, University of Warwick, Coventry, CV4 7AL, United Kingdom}

\ead{luca.garzotti@ccfe.ac.uk}

\begin{abstract}
Reactor grade plasmas are likely to be fuelled by pellet injection. This technique transiently perturbs the profiles, driving
the density profile hollow and flattening the edge temperature profile. After the pellet perturbation, the density and temperature profiles 
relax towards their quasi-steady-state shape. Microinstabilities influence plasma confinement and will play a role in determining the evolution 
of the profiles in pellet fuelled plasmas. In this paper we present the microstability analysis of pellet fuelled H-mode MAST plasmas. Taking 
advantage of the unique capabilities of the MAST Thomson scattering system and the possibility of synchronizing the eight lasers with the pellet 
injection, we were able to measure the evolution of the post-pellet electron density and temperature profiles with high temporal and spatial 
resolution. These profiles, together with ion temperature profiles measured using a charge exchange diagnostic, were used to produce
equilibria suitable for microstability analysis of the equilibrium changes induced by pellet injection. This analysis, carried out using the local 
gyrokinetic code GS2, reveals that the microstability properties are extremely sensitive to the rapid and large transient excursions of 
the density and temperature profiles, which also change collisionality and $\beta_e$ significantly in the region most strongly affected 
by the pellet ablation.
\end{abstract}

%Uncomment for PACS numbers title message
% Uncomment for Submitted to journal title message
\submitto{\PPCF}
% Comment out if separate title page not required
\maketitle

\section{Introduction}\label{introduction}
Reactor grade plasmas are likely to be fuelled by pellet injection. One of the main differences between this technique and fuelling
by gas puffing is that the density and temperature profiles are not stationary, but undergo sudden perturbations on every pellet injection 
and then relax to the quasi-steady-state profile shapes. In particular, the density profile becomes hollow and develops steeper gradients to 
either side of the radius with maximum pellet ablation and the temperature profile flattens.

The evolution of the kinetic profiles following the injection of a pellet might be expected to affect the microstability and 
the transport properties of the plasma, particularly in reactor relevant scenarios where transport inducing MHD phenomena, like, for
example, neoclassical tearing modes (NTMs) or edge localized modes (ELMs), are suppressed or greatly mitigated. Therefore, it is interesting 
to perform a detailed microstability analysis of the experimental profiles measured after pellet injection.

In this paper we take advantage of the unique capabilities of the MAST Thomson scattering system \cite{Thomson} and the possibility of 
synchronizing the eight lasers with the pellet injection to measure the evolution of the post-pellet electron density and temperature profiles 
with high temporal and spatial resolution and use the experimental profiles to generate equilibria that are suitable for microstability analysis. 
The local gyrokinetic code GS2 \cite{Kotschenreuther95} has been used to study the predicted evolution of the microstability following the pellet 
injection.

This paper is organized as follows. In section \ref{expsetup} we describe the MAST pellet injector, the Thomson scattering system
and the reconstruction of the equilibria used for microstability analysis. Then, in section \ref{microanalysis}, we describe 
the results from local gyrokinetic simulations and section \ref{discussion} discusses the possible implication of these results 
for transport in pellet fuelled plasmas. Final conclusions are drawn in section \ref{conclusions}.

\section{Experimental set-up}\label{expsetup}
The experiments analysed in this study were performed using the MAST pellet injection system \cite{Ribeiro01}. 
The MAST pellet injector is a pneumatic injector capable of firing pellets with mass $0.6 \cdot 10^{20}$, $1.2 \cdot 10^{20}$ and 
$2.4 \cdot 10^{20}$ deuterium atoms at speeds between 250 and 400 m/s. The injector is situated 20 m from the tokamak and the pellets are 
delivered through a guide tube and injected  vertically from the high field side of the plasma.

\begin{figure}[htbp]
\begin{center}
\includegraphics[width=6cm]{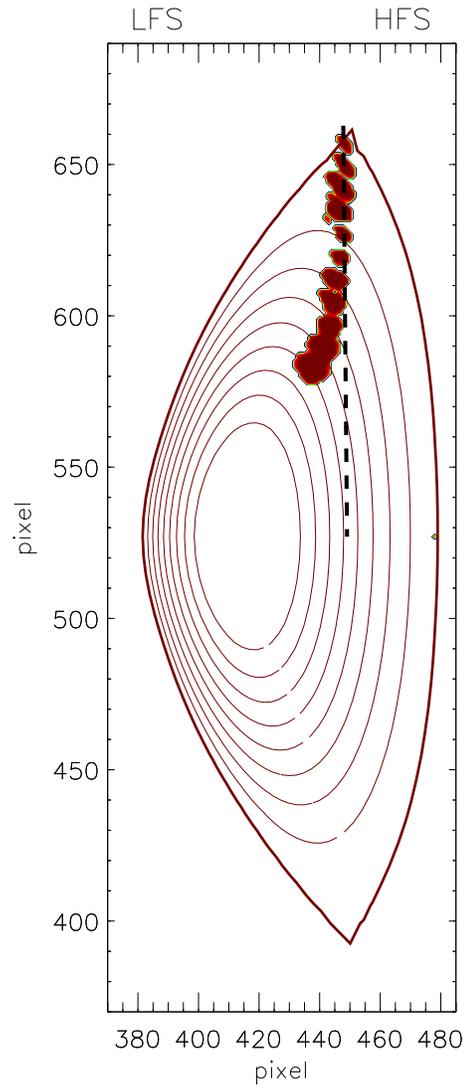}
\caption{\label{cross}Fast camera image of the pellet trajectory for MAST shot 24743. The picture is obtained
by superimposing images of the pellet cloud taken at 0.2 ms intervals. On the camera image are also projected a 
contour map of the poloidal flux and the pellet injection line.}
\end{center}
\end{figure}

Figure \ref{cross} shows the trajectory of a MAST pellet as seen by a fast camera installed on the plasma equatorial plane and looking tangentially 
at the pellet injection poloidal plane. A contour map of the poloidal flux $\Psi$ and the pellet injection line are also projected onto the camera image. 
In this particular case (MAST shot 24743, plasma current $I_p=900$ kA, toroidal magnetic field $B_T=0.54$ T, additional neutral beam heating $P_{NBI}=3.5$ MW 
and on-axis electron density and temperature $n_{e0} \approx 4.5 \cdot 10^{19}$ m$^{-3}$ and $T_{e0} \approx 1.0$ keV) the target plasma was in double null 
divertor configuration with geometrical major radius $R=0.79$ m, minor radius $a=0.59$ m, elongation $\kappa=1.92$ and triangularity $\delta=0.43$. 
The pellet speed was $\sim 300$ m/s and the pellet mass (derived from the density increase) was $\sim 1.9 \cdot 10^{20}$ deuterium atoms.

The electron density and temperature profiles used for the analysis were measured with the MAST Thomson scattering system at 130 points 
on the equatorial plane of the machine, covering  the outboard and inboard side of the plasma and with spatial resolution of $\approx$1 cm. 
The system consists of eight neodymium-doped yttrium aluminum garnet (Nd:YAG) lasers each one of which can fire with a repetition rate of 30 Hz. 
The lasers can be fired at regular intervals providing a time resolution of 240 Hz (regular mode) or in rapid succession to analyze fast phenomena of the 
duration of typically few ms (burst mode). For our study, the Thomson scattering system was operated in burst mode. A system based on a field-programmable 
gate array (FPGA) provides a smart trigger that stops the sequence of laser bursts when the pellet leaves the injector and restarts it when the pellet enters 
the plasma. The signals to stop and restart the lasers bursts are provided by optical barriers located at the beginning and at the end of the launch tube. 
In this way we can guarantee the synchronization of the measurement of the density and temperature profiles with the pellet ablation and the subsequent 
relaxation of the profiles. Typical spacing between the lasers was 1.5 ms. The ion temperature was measured by charge exchange spectroscopy with temporal 
resolution of 5 ms and spatial resolution of 1 cm. The profiles were linearly interpolated between neighbouring time slices. Typical average errors for 
the shot analysed in this paper are less than 5\% for the electron density and temperature and between 5\% and 10\% for the ion temperature.

The error bars on the density and temperature gradients are more difficult to estimate. By varying the parameters of the fits, we 
verified empirically that the quality of the experimental measurements can constrain the gradients to within 25\%. Test runs of GS2 where the 
gradients were independently increased or decreased by this amount did not significantly change the results. We note also that these errors are 
much smaller than the dramatic changes in $R/L_{n_e}$ and $R/L_{T_{e,i}}$ during the pellet transient (as shown in table \ref{GS2}).

Figure \ref{timing} shows the electron density and temperature profiles and the relative timing of the Thomson scattering measurements with respect to the line 
integrated density measured by the MAST interferometer for MAST shot 24743. It can be seen that there is no pre-pellet measurement for this shot, as the first 
measurements is taken at $t=314.8$ ms, corresponding to the beginning of the ablation just after the pellet entered the plasma. Therefore, we used MAST shot 24541 
(identical to 24743) as a no-pellet reference case and the profiles at $t=317.0$ ms for shot 24541 were taken as representative of pre-pellet profiles. For the pellet 
shot profiles, the time slices selected for the microstability analysis are: $t=317.8$ ms and $t=323.8$ ms.

The first time slice from the pellet discharge is taken immediately after the end of the pellet ablation and the second is taken towards the end of the post-pellet 
relaxation of the density and temperature profiles. In relative terms, we will label the time slice from the reference discharge with $t_0$, the first time slice 
from the pellet discharge with $t_0+t_{abl}$ and the second time slice from the pellet discharge with $t_0+10$ ms.

\begin{figure}[htbp]
\begin{center}
\includegraphics[width=15cm]{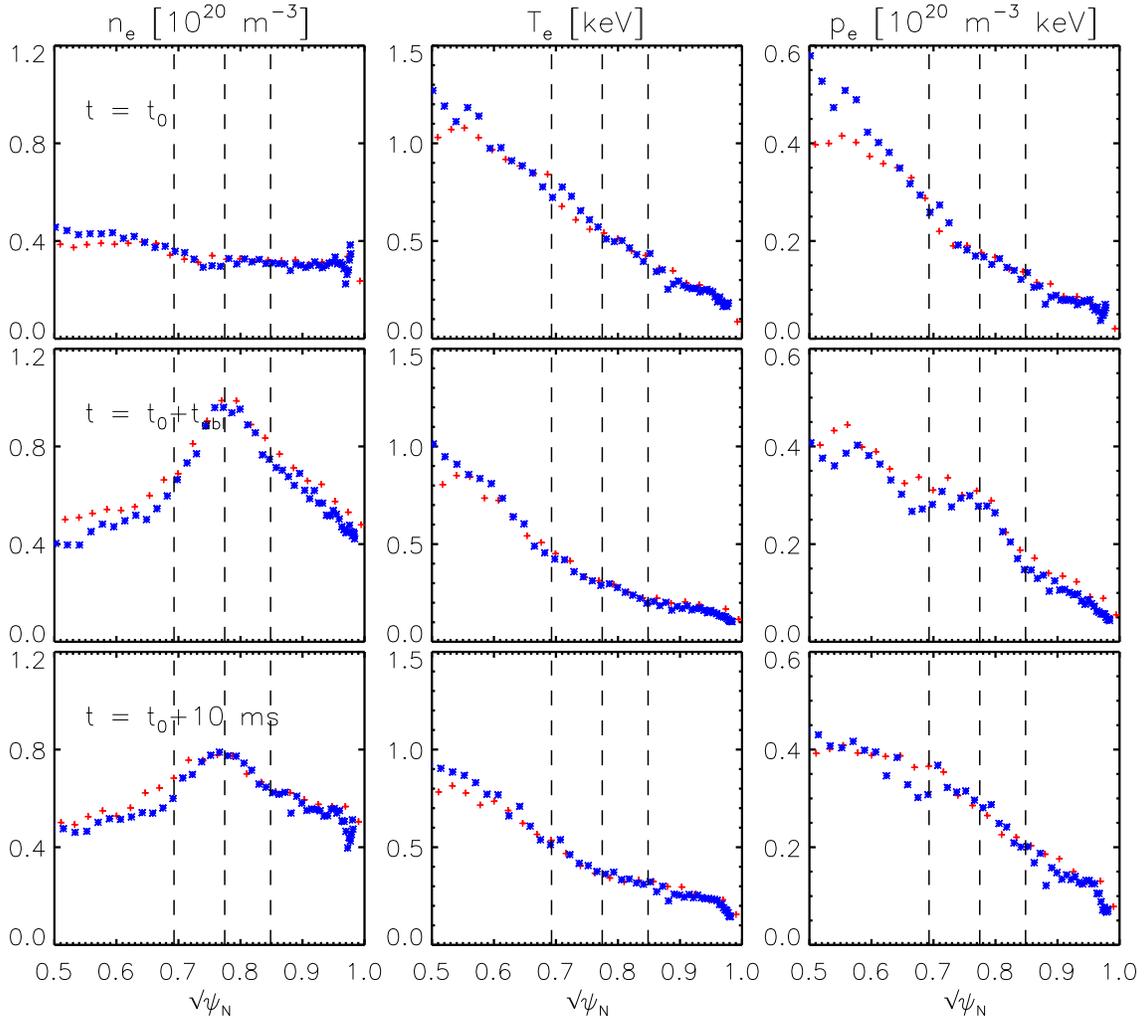}
\caption{\label{exthomson}Electron density (left), temperature (middle) and pressure (right) profiles selected for the 
microstability analysis described in the main text. The first time slice is taken from shot 24541 (no-pellet reference) and the 
second and third time slices are taken from shot 24743 (pellet shot). Blue asterisks indicate high field side profiles and red 
crosses indicate low field side profiles. The vertical dashed lines indicate the flux surfaces selected for the microstability
analysis.}
\end{center}
\end{figure}

Figure \ref{exthomson} shows the time sequence of the selected electron density, temperature and pressure profiles. The profiles are plotted as a function of the square 
root of the normalised poloidal flux $\sqrt{\psi_N}$ and the low field side and high field side part of the profiles are indicated by red crosses and blue asterisks 
respectively. It can be seen that the profiles are not completely inboard/outboard symmetric and therefore for the microstability analysis a procedure of averaging 
and smoothing was applied. 

\begin{figure}[htbp]
\begin{center}
\includegraphics[width=12cm]{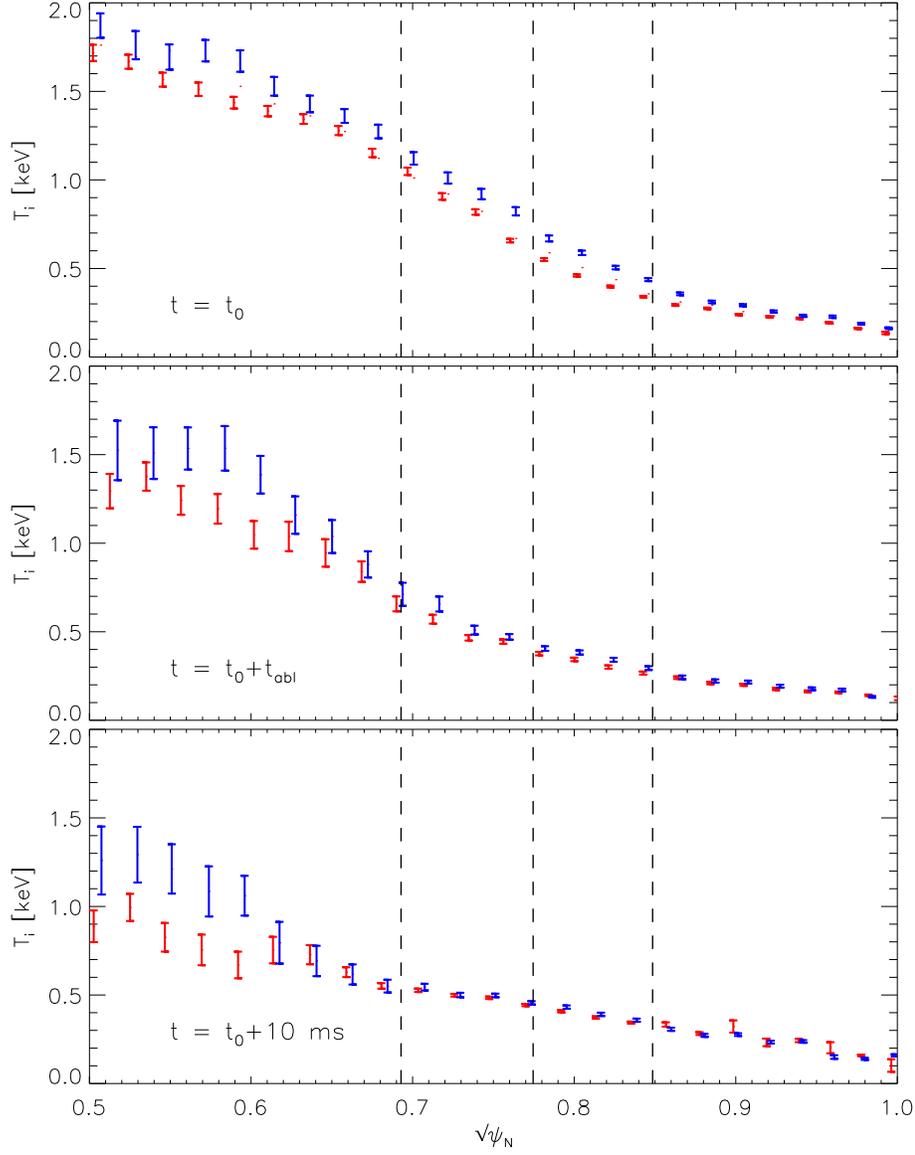}
\caption{\label{excx}Ion temperature profiles selected for the microstability analysis described in the main text. 
The first time slice is taken from shot 24541 (no-pellet reference) and the second and third time slices are taken from shot 24743 
(pellet shot). The red and blue points correspond to the two ion temperature profiles obtained from the two beam lines available on MAST.
The vertical dashed lines indicate the flux surfaces selected for the microstability analysis.}
\end{center}
\end{figure}

Figure \ref{exthomson} also indicates the three radial locations selected for the microstability analysis, corresponding to $\sqrt{\psi_N}=0.69$, $\sqrt{\psi_N}=0.77$ 
and $\sqrt{\psi_N}=0.85$. These three flux surfaces are representative of: the region inside the peak of the pellet ablation profile, where the density gradient is positive 
and the temperature gradient is steeper; the region around the peak of the pellet ablation profile, where the density gradient is almost zero; and  the 
region outside the peak of the pellet ablation profile, where the density gradient is negative (but steeper than in the case without pellet) and 
the temperature gradient is flatter. Figure \ref{excx} shows the ion temperature profiles for the three time slices selected for the microstability analysis

The equilibria for microstability analysis using GS2 were obtained from TRANSP analysis \cite{TRANSP}. As 
mentioned previously, the density and temperature profiles were averaged between high field side and low field side and the resulting average 
was smoothed using spline fits. Equilibrium reconstructions were constrained to match both the position of the plasma edge as measured by a 
$D_{\alpha}$ emission linear array and the position of the magnetic axis obtained using motional Stark effect measurements. 
Anomalous fast ion diffusion was used in TRANSP to obtain model fast ion density profiles that were compatible with the measured 
neutron yield.

Equilibrium parameters for the three surfaces where local microstability analysis was performed are summarized in table \ref{GS2}. It is worth noting that 
the transient excursions of $R/L_{n_{e}}$ and $R/L_{T_{e,i}}$ induced by the pellet are much larger than the errors in these quantities.

\begin{table}
\caption{\label{GS2} Local equilibrium quantities on the three flux surfaces and at the three
times of interest, which were used as inputs to GS2. $n_e$ is the electron density, 
$T_e$ and $T_i$ are the electron and ion temperature, $R/L_{n_e}$ is the normalized density 
gradient, $R/L_{T_e}$ and $R/L_{T_i}$ are the normalized electron and ion temperature gradients,
$\eta_e=L_{n_e}/L_{T_e}$, $\eta_i=L_{n_e}/L_{T_i}$, $\nu_e$ is the electron collision frequency 
normalised to $v_{ti}/a$, where $v_{ti}$ is the thermal ion velocity and $a$ is the minor radius, 
and $\beta_e$ is the electron $\beta$.}
%\begin{indented}
%\item[]
\begin{tabular}{@{}clcccccccccc}
\br
&&$n_e$&$T_e$&$T_i$&$R/L_{n_e}$&$R/L_{T_e}$&$\eta_e$&$R/L_{T_i}$&$\eta_i$&$\nu_e$&$\beta_e$\\
&&[10$^{19}$ m$^{-3}$]&[keV]&[keV]& & &&&&&[\%]\\
\mr
$dn/dr>0$ &$t_0$&3.29&0.70&1.02&1.47&6.25&4.25&5.95&4.05&0.15&4.4\\
($\sqrt{\psi_N} = 0.69$) &$t_0+t_{abl}$&5.67&0.50&0.76&-7.47&6.95&-0.93&8.26&-1.11&0.46&5.6\\
&$t_0+10$ ms&5.86&0.56&0.66&-3.93&5.73&-1.46&5.45&-1.39&0.43&5.1\\
\mr
$dn/dr \approx 0$&$t_0$&3.00&0.46&0.64&1.18&5.83&4.94&7.48&6.34&0.31&2.5\\
($\sqrt{\psi_N} = 0.77$)&$t_0+t_{abl}$&8.51&0.30&0.40&0.73&7.01&9.60&8.47&11.60&1.94&4.5\\
&$t_0+10$ ms&6.70&0.37&0.45&0.39&4.92&12.61&4.74&12.25&1.08&4.0\\
\mr
$dn/dr<0$&$t_0$&2.83&0.34&0.44&1.06&5.01&4.72&5.05&4.76&0.54&1.6\\
($\sqrt{\psi_N} = 0.85$)&$t_0+t_{abl}$&6.73&0.216&0.31&4.75&4.97&1.05&0.82&0.17&2.86&2.7\\
&$t_0+10$ ms&5.91&0.30&0.38&3.31&3.37&1.02&1.03&0.31&1.41&3.0\\
\br
\end{tabular}%
%\end{indented}
\end{table}

\section{Microstability analysis}\label{microanalysis}

Microstability analysis was performed with the local gyrokinetic code GS2 \cite{Kotschenreuther95}, which solves the gyrokinetic Vlasov/Maxwell
system of equations in toroidal plasmas for an arbitrary number of plasma species in reduced volume flux-tube geometry that is aligned with the equilibrium 
magnetic field. GS2 includes electromagnetic perturbations and collisions, which are described by means of an energy and pitch angle scattering operator.
In the runs performed for the purpose of this study, GS2 was run with five species (electrons, D ions, H ions, fast D ions and C ions). The runs were 
linear and electromagnetic, using only pitch angle scattering in the collision operator model. In this first analysis we have neglected the contributions 
of sheared equilibrium flows. However, estimates of the $E \times B$ shear on MAST seem to indicate that it could suppress turbulence with very 
long wavelengths ($k_y \rho_i \sim 0.1$) but not the turbulence at $k_y \rho_i \sim 1.0$ typical of TEMs, which are associated with particle transport
\cite{Roach09,Field11,Henderson13}. 

The local gyrokinetic treatment is most reasonable when the separation between rational surfaces is much less than the equilibrium gradient
scale length. The most interesting modes in this paper that cause particle transport are TEMs, which are most unstable at $k_y \rho_i \sim 1$ where this
condition is very well satisfied. The local approximation, however, may be less reliable at very low $k_y \rho_i$ and on surfaces with low magnetic shear.

The main results from the microstability analysis performed with GS2 are presented in figures \ref{GS2res1}, \ref{GS2res2} and \ref{GS2res3} 
where the growth rates and the real frequencies of the dominant unstable modes are plotted as a function of $k_y \rho_i$ for the three time slices 
at each of the three flux surfaces chosen for the analysis. These GS2 results indicate that the post-pellet microstability is changed dramatically by 
the rapid and large excursions of density and temperature profiles, collisionality and $\beta_e$. The response of the plasma microstability to these 
variations is complex and strongly depends on radial position with respect to the pellet deposition peak. The dominant modes differ on the three surfaces 
analysed, and their growth rates evolve differently during the transient profile changes.

In the following subsections we will describe the results of the microstability analysis for each of the flux surfaces considered in this study.

\subsection{$dn/dr>0$ ($\sqrt{\psi_N} = 0.69$)}
On the inner surface, where $dn/dr>0$ ($\sqrt{\psi_N} = 0.69$), figure \ref{GS2res1} shows that the dominant modes, at all $k_y \rho_i$ values, 
are progressively stabilized by pellet injection. This is due to the greatly increased density gradient, which reduces
both $\eta_i=L_{n_e}/L_{T_i}$ and $\eta_e=L_{n_e}/L_{T_e}$. The reduction in $\eta_i$ is responsible for the stabilization of ion temperature gradient 
(ITG) modes in the low/intermediate $k_y \rho_i$ region of the spectrum ($k_y \rho_i < 1$)and the reduction in $\eta_e$ 
stabilizes electron temperature gradient (ETG) modes at the high $k_y \rho_i$ region of the spectrum ( $k_y \rho_i > 1$ ).

\subsection{$dn/dr\approx 0$ ($\sqrt{\psi_N}=  0.77$)}
The microinstability evolution is very different at the middle surface at the peak of the pellet ablation profile, where $dn/dr\approx 0$ 
($\sqrt{\psi_N}=  0.77$). Immediately after pellet injection, we observe an initial decrease in the  growth rate of modes with $k_y \rho_i<0.6$ 
and a simultaneous increase of the growth rates of modes with $0.6<k_y \rho_i<2$. The perpendicular wavenumber of the most unstable mode in the 
region $k_y \rho_i < 1$ moves from $k_y \rho_i \approx 0.3$ to $k_y \rho_i \approx 0.9$. Towards the end of the profile relaxation phase, the 
growth rates of the dominant modes with $0.6<k_y \rho_i<0.8$ return close to their pre-pellet values, 
and a stability window appears in the wavenumber region $0.8<k_y \rho_i<2$. 
Modes with $k_y \rho_i>2$ are only very weakly affected by the pellet injection.

\begin{figure}[htbp]
\begin{center}
\includegraphics[width=12cm,angle=270]{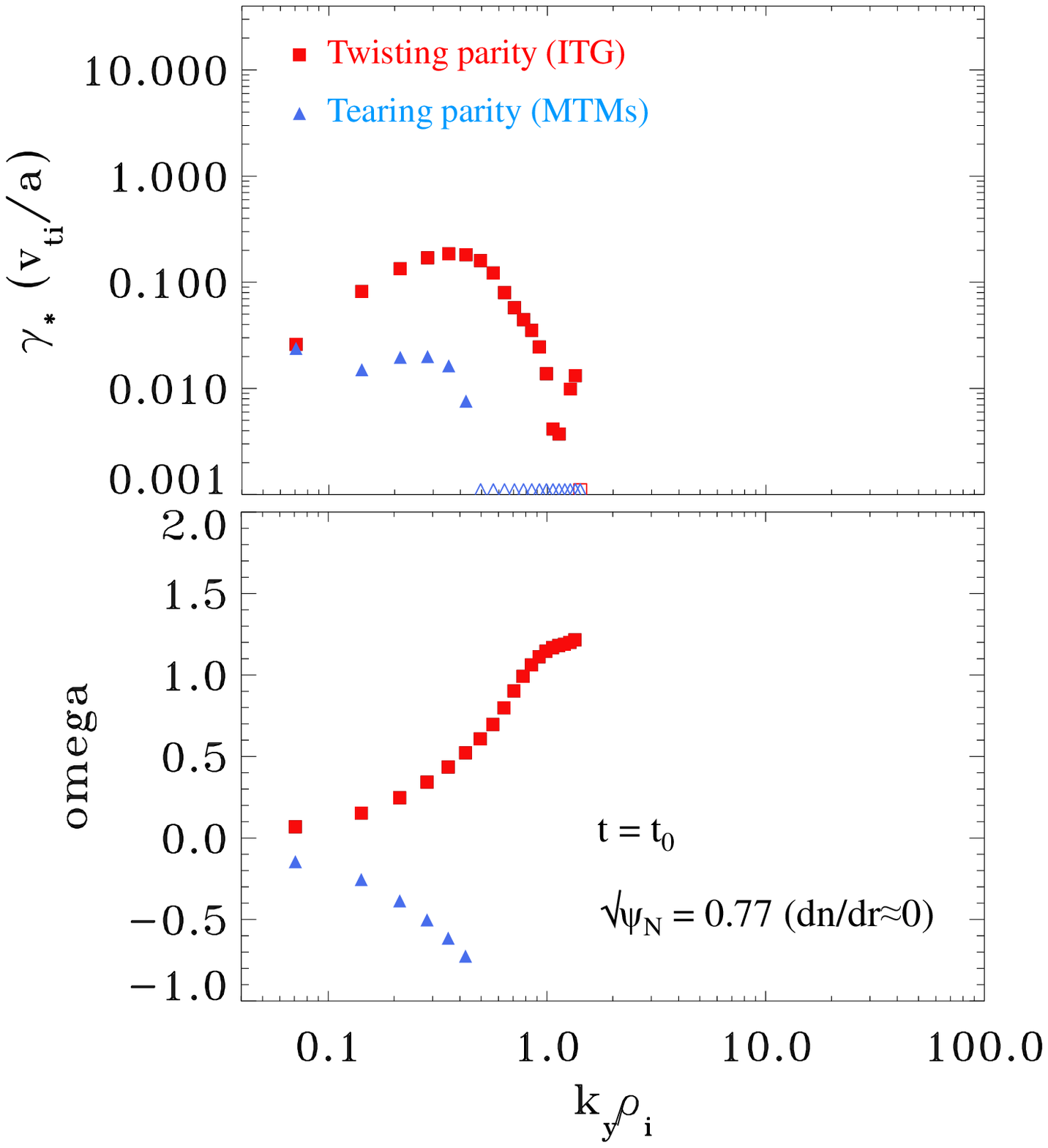}
\caption{\label{microtearingpre}Results of the GS2 microstability analysis at $\sqrt{\psi_N}=  0.77$ ($dn/dr \approx 0$) and $t=t_0$.
Red squares show the growth rate of modes with twisting parity, identified as ITG modes driven by the ion temperature gradient.
Blue triangles show the growth rate of modes with tearing parity, identified as subdominant MTMs. Top: growth rates, bottom: real frequencies. 
Both quantities are normalized to $v_{ti}/a$ (where $v_{ti}$ is the ion thermal velocity and $a$ is the plasma minor radius).
Full symbols correspond to unstable modes and open symbols to stable modes.}
\end{center}
\end{figure}

The interpretation of these results is more complicated than for the inner surface. At $t=t_0$, the dominant modes with $k_y \rho_i<2$
have twisting parity and are ITG modes that are largely driven by the ion temperature gradient\footnote{GS2 simulations with a 
Boltzmann electron response give similar results for the dominant modes in this region of the spectrum, demonstrating that electron 
physics has little role in driving these modes.}. We have also determined the growth rates of the fastest growing tearing parity modes in the 
same region of the spectrum, by using a parity filter that exploits the up-down symmetry of the adopted equilibrium model. Figure 
\ref{microtearingpre} shows the resulting growth rates and reveals that at $t=t_0$ microtearing modes (MTMs) are also unstable, but 
that they are subdominant to ITG modes at this initial time. 

\begin{figure}[htbp]
\begin{center}
\includegraphics[width=12cm,angle=270]{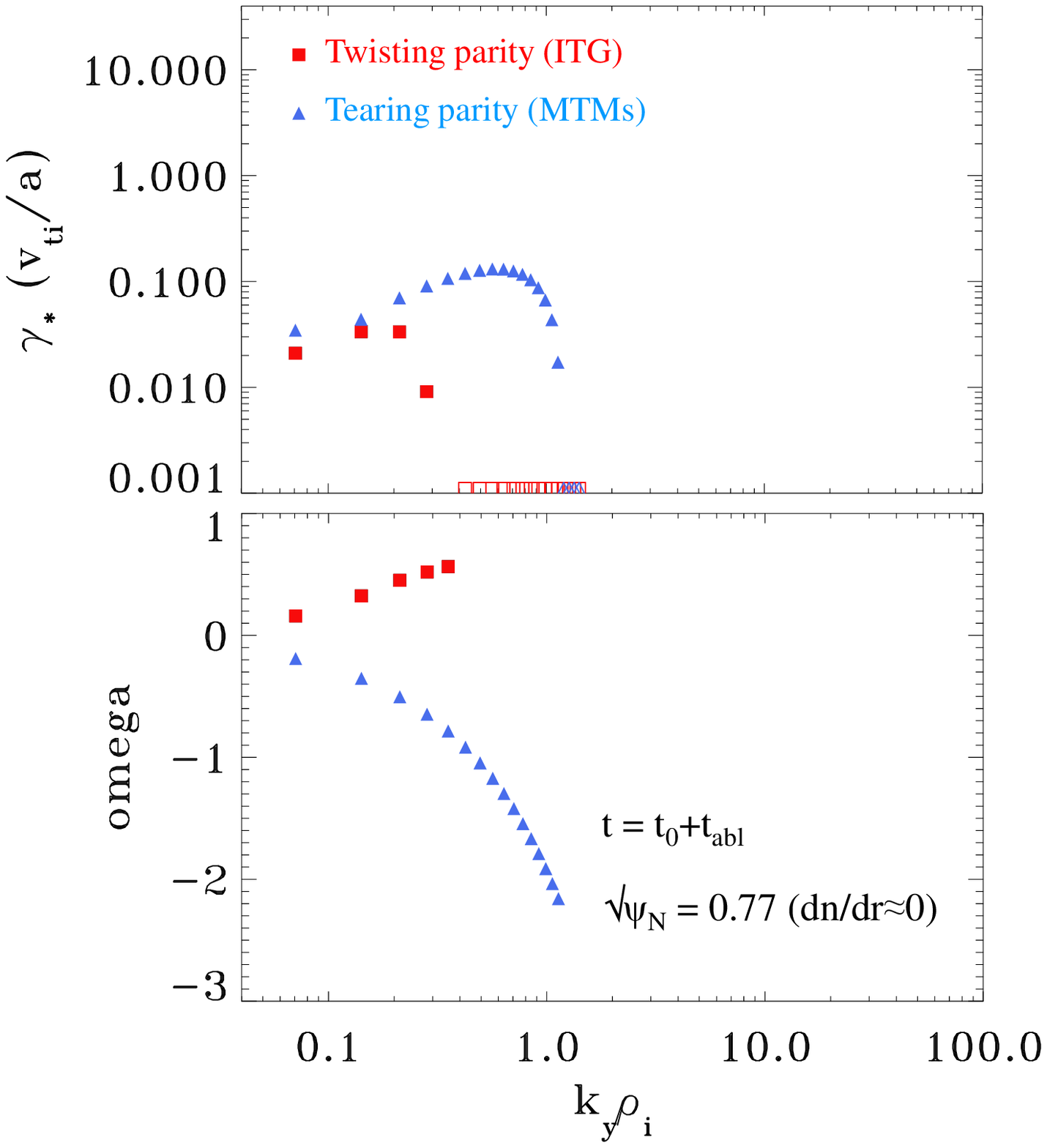}
\caption{\label{microtearingpost}Results of the GS2 microstability analysis at $\sqrt{\psi_N}=  0.77$ ($dn/dr \approx 0$) and $t=t_0+t_{abl}$,
showing evidence of a mode switch between ITG modes and MTMs. Red squares show the growth rate of modes with twisting parity, identified as 
subdominant ITG modes. Blue triangles show the growth rate of modes with tearing parity, identified as MTMs driven by the increased $\beta_e$.
Top: growth rates. Bottom: real frequencies. Both quantities are normalized to $v_{ti}/a$ (where $v_{ti}$ is the ion thermal velocity and $a$ 
is the plasma minor radius). Full symbols correspond to unstable modes and open symbols to stable modes.}
\end{center}
\end{figure}

Immediately after pellet injection, $\beta_e$, $R/L_{Te}$ and electron collisionality increase significantly, 
which suppresses the ITG growth rates and simultaneously destabilises MTMs \cite{Applegate04, Applegate07}.
At $t=t_0 + t_{abl}$ the increasingly unstable MTMs take over as the dominant modes in this region of the spectrum. 
Figure \ref{microtearingpost} shows the growth rates of the dominant modes at $t=t_0+t_{abl}$ (all with tearing parity), 
together with the growth rates of the fastest growing twisting parity modes (ITG), which are then sub-dominant to MTMs. 
Increasing $\beta_e$ during the H-mode pedestal recovery in MAST has similarly been shown to enhance the growth rates of MTMs
that dominate in the edge plateau plasma \cite{Dickinson2012}. MTM growth rates are enhanced at large trapped 
particle fraction \cite{Dickinson2013}, so that the impact of pellet induced MTMs might be expected to depend 
strongly on the radius of maximum pellet deposition.

Towards the end of the relaxation of the pellet deposition profile ($t=t_0+10$ ms), when collisionality, $R/L_{Te}$ and $\beta_e$ 
decrease, the MTM growth rates subside slightly. At $t=t_0+10$ ms a window of stability has opened up in the region $0.8 < k_y \rho_i < 2$. 
The opening up of a similar stability window, but between dominant ITG/TEM modes at low $k_y\rho_i$ and ETG modes at high $k_y \rho_i$, 
has been explained in terms of the faster detrapping of electrons at higher electron collision frequency \cite{Roach09}.
In the high $k_y \rho_i$ region of the spectrum, it can be seen that ETG modes are hardly affected by the pellet injection because 
at the peak of the density profile there is no increase in density gradient and therefore no reduction in $\eta_e$.

\subsection{$dn/dr<0$ ($\sqrt{\psi_N}=  0.85$)}
At the outer surface with $dn/dr<0$ ($\sqrt{\psi_N}=  0.85$), after the pellet ablation, we see a decrease in the growth rates of the dominant modes 
at $k_y \rho_i<0.8$, an increase in the growth rates of modes at $0.8<k_y \rho_i<2$ and complete stabilization of modes at $k_y \rho_i>2$.
In the low/intermediate $k_y \rho_i$ region of the spectrum, we can see a shift in $k_y \rho_i$ of the most unstable mode from $\sim 0.3$ to 
$\sim 1.0$. Towards the end of the profile relaxation phase, the growth rates of modes with $0.8<k_y \rho_i<2$ fall slightly and modes with 
$k_y \rho_i>2$ remain stable.

A series of tests similar to those performed at the middle flux surface ($\sqrt{\psi_N}=  0.77$) have allowed us to establish that, before pellet
injection, the region $k_y \rho_i<1.0$ is dominated by ITG modes driven predominantly by the ion temperature gradient \footnote {This is supported by 
the similar growth rate spectrum, in this wavenumber range, obtained from calculations that assume an electron Boltzmann response.}. Moreover, at $t=t_0$, 
modes with tearing parity are also unstable at $k_y \rho_i<0.3$, but subdominant to ITG modes. 

After pellet injection, at $t=t_0 + 10$ ms, the ITG modes become stabilized by the increased density gradient and the decreased ion temperature gradient, 
both leading to a decreased $\eta_i$, leaving MTMs dominating in the region of the spectrum $k_y \rho_i<0.4$, albeit with low growth rates. The modes with 
$0.4<k_y \rho_i<2$ are identified as trapped electron modes (TEMs), as they are fully stabilized when the trapped electron drive is removed (e. g. assuming Boltzmann 
electrons). Independently varying the gradients of density, and of ion and electron temperature, revealed the TEMs at $0.4<k_y \rho_i<2$ are predominantly driven 
by the electron density gradient and are driven more unstable after pellet injection by the increase in $R/L_{n_e}$. At the same time the increased electron density 
gradient reduces $\eta_e$ and stabilizes ETG modes with $k_y \rho_i>2$, as for the surface at $\sqrt{\psi_N} = 0.69$.

The impacts of the pellet enhanced density gradients on TEMs are in stark contrast inboard and outboard of the peak pellet ablation 
surface. On the outer surface the enhanced density gradient drives TEMs more {\em unstable}, but on the inner surface, where the density gradient 
is strongly {\em reversed}, the TEMs are {\em stabilised}. Artificially reversing magnetic drifts in the gyrokinetic analysis of the 
post-pellet inner surface recovers strongly unstable TEMs, suggesting that TEMs are stabilised by favourable drifts on this surface. 
This is unsurprising, as it is well known that collisionless TEM (CTEM) growth rates are sensitive to the precession drift of trapped 
electrons \cite{Roach95}, and that CTEMs can be stabilised when the electron diamagnetic drift opposes the electron precession 
drift \cite{Tang78}.

While reversal of the density gradient makes magnetic drifts {\em more favourable} in the outboard region near $\theta=0$, the drifts 
become {\em less favourable} in the conventionally stable inboard plasma near $\theta=\pi$. To investigate this further, we ran collisionless 
gyrokinetic simulations to obtain, for the post-pellet inner surface, the CTEM growth rate, $\gamma_{CTEM}$, as a function of
ballooning angle, $\theta_0$. With drifts in the experimental direction, CTEMs are unstable and $\gamma_{CTEM}$ depends weakly on $\theta_0$, but 
with drifts artificially reversed the unstable CTEMs are highly ballooning with $\gamma_{CTEM}$ strongly peaked around $\theta_0 = \pi$. It is significant 
to note that {\em without collisions} CTEMs would be {\em unstable} on the post-pellet inner surface. {\em With the experimental 
level of electron collisions}, however, {\em TEMs are stabilized} on the post-pellet inner surface in MAST. This suggests that in lower 
collisionality plasmas the post-pellet inner surface may be more unstable to TEMs, which could improve the penetration of pellet particles into the core.

\subsection{Comparison with previous analysis}
We have repeated the microstability analysis of an older MAST pellet shot presented in \cite{Valovic08}, where density and temperature profiles were measured 
with the previous Thomson scattering system, which had a lower temporal and spatial resolution than the current system. In that shot the pellet mass was 
0.7$\cdot$10$^{20}$ deuterium atoms (i. e. $\sim 60$\% smaller than the pellet considered in the present study) and the analysis was limited to one radial flux 
surface (corresponding to the maximum $R/L_{T_{e}}$ surface inside the pellet ablation peak) at one time slice (5 ms after pellet injection) and considered 
only the region of the spectrum $0.08 < k_y \rho_i< 0.5$. 

Our repeat analysis recovers the increase in the growth rates of modes in the spectral range $0.08 < k_y \rho_i < 0.3$, a trend opposite to the decrease in growth 
rates in this region of the $k_y \rho_i$ spectrum that we report here from our analysis of the inner surface in discharge 24743. This difference is due to a shallower 
density gradient inside the pellet ablation peak in the older discharge, which resulted in higher $\eta_e$ and $\eta_i$ values. This indicates that the microstability in 
the region of positive density gradient depends on pellet size. It is worth noting that for ITER the relative pellet size will be smaller than in present day experiments. 
This test demonstrates the high sensitivity of the microstability analysis to the input parameters and the need for high resolution measurements to make reliable predictions.

\section{Discussion}\label{discussion}
The results of the microstability analysis presented in this paper predict an asymmetric evolution of the post-pellet density and temperature profiles. 
If extrapolated to reactor grade plasma, they would imply that, if the density gradient inside the peak of the pellet deposition profile is too steep,
microturbulence in this region with positive density gradient may be stabilized and the anomalous diffusion of the pellet particles towards 
the plasma core due to microinstabilities would be reduced. For the pellet fuel to reach the plasma core, one would have to rely on other mechanisms like
fast transport associated with the $\nabla B$ drift of the pellet cloud or larger scale MHD instabilities as described for example in \cite{Annibaldi04}.

\begin{figure}[htbp]
\begin{center}
\includegraphics[width=12cm,angle=270]{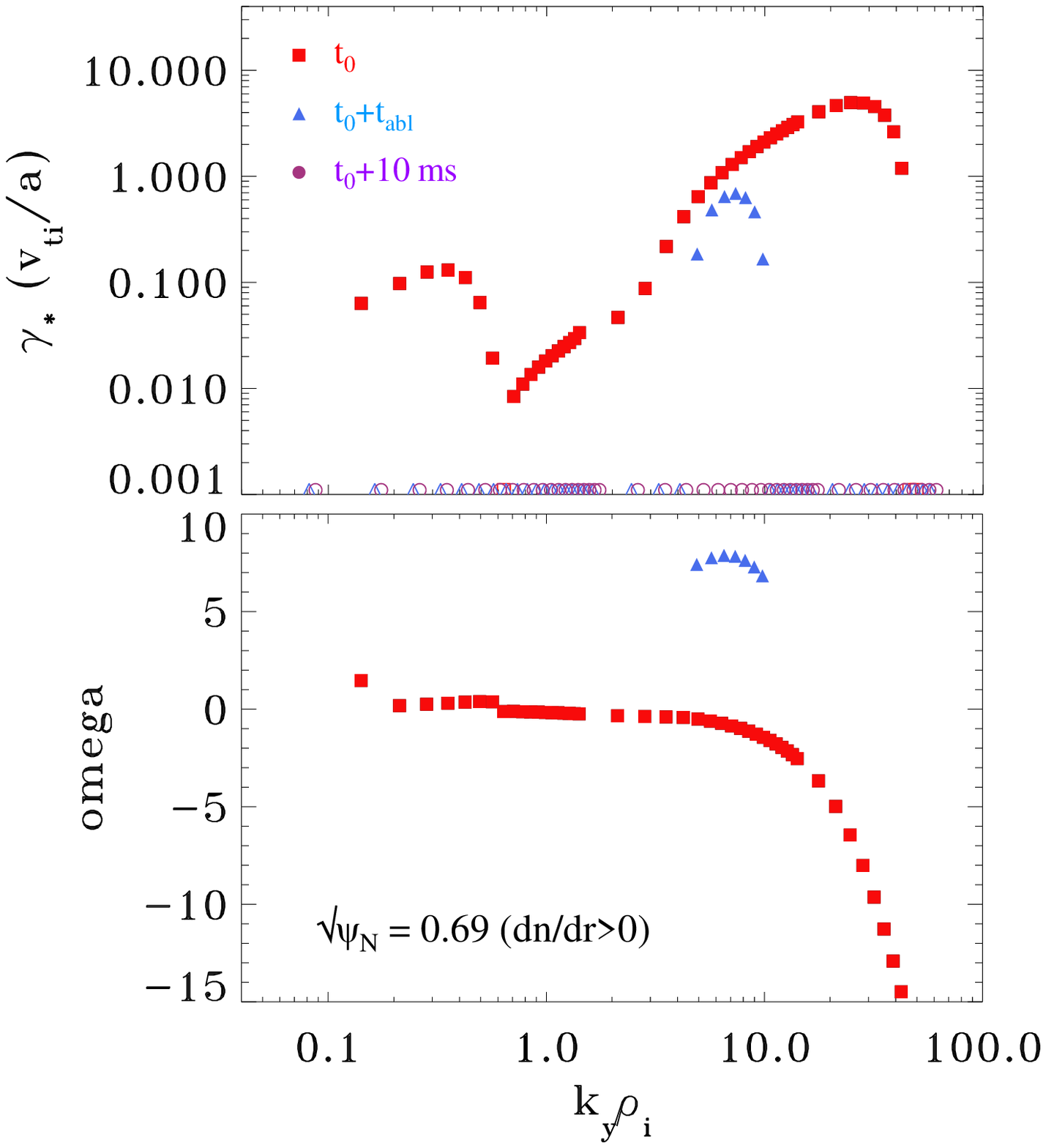}
\caption{\label{GS2res1}Results of the GS2 microstability analysis at $\sqrt{\psi_N} =  0.69$ ($dn/dr>0$). Top: growth rates, 
bottom: real frequencies. Both quantities are  normalized to $v_{ti}/a$ (where $v_{ti}$ is the ion thermal velocity and $a$ is 
the plasma minor radius). Full symbols correspond to unstable modes and open symbols to stable modes.}
\end{center}
\end{figure}

\begin{figure}[htbp]
\begin{center}
\includegraphics[width=12cm,angle=270]{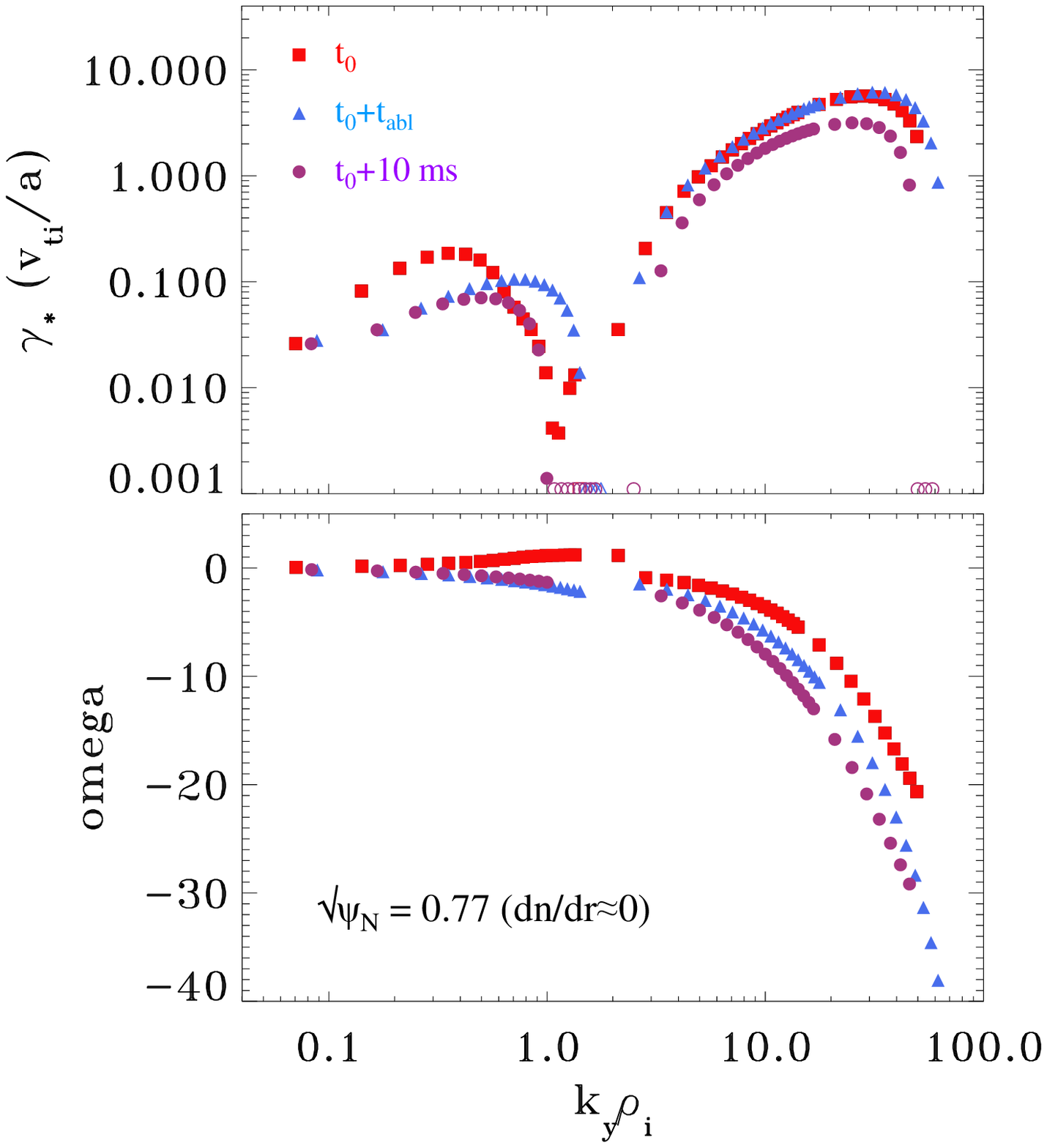}
\caption{\label{GS2res2}Results of the GS2 microstability analysis at $\sqrt{\psi_N}=  0.77$ ($dn/dr\approx 0$).
Symbol convention as in figure \ref{GS2res1}.}
\end{center}
\end{figure}

\begin{figure}[htbp]
\begin{center}
\includegraphics[width=12cm,angle=270]{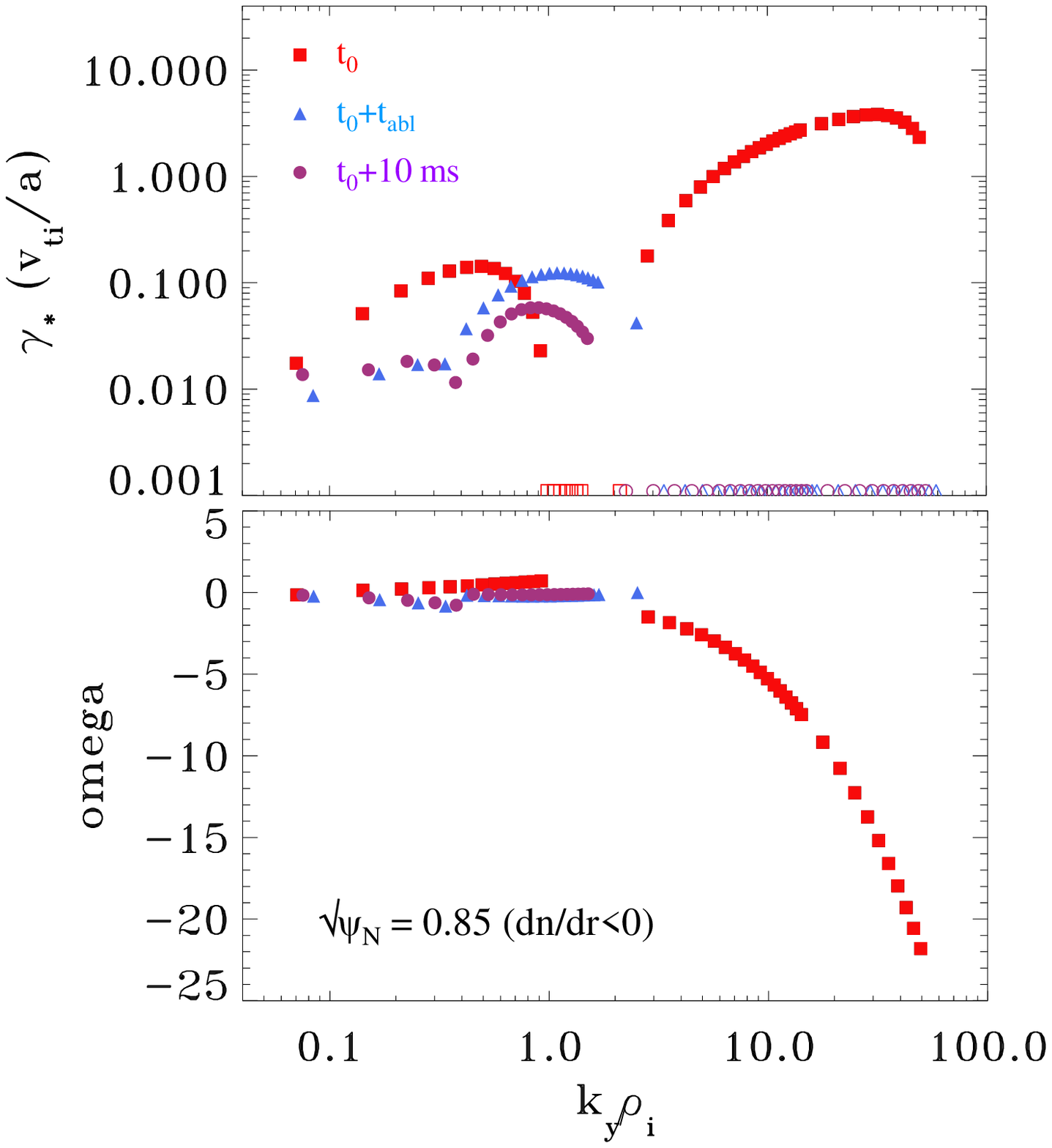}
\caption{\label{GS2res3}Results of the GS2 microstability analysis at $\sqrt{\psi_N}=  0.85$ ($dn/dr<0$).
Symbol convention as in figure \ref{GS2res1}.}
\end{center}
\end{figure}

Computing transport coefficients from gyrokinetic simulations during the post-pellet profile evolution is challenging. The gyrokinetic turbulence evolves 
over the timescale corresponding to the linear growth time (of the order of 20 $\mu$s for the ITG peak at $k_y \rho_i \sim 0.4$ in figure \ref{GS2res1}), 
which is much shorter than the post-pellet profile evolution timescale (typically of the order of  several milliseconds). In this study we have performed only 
linear local gyrokinetic simulations and we have not attempted to obtain detailed transport coefficients from our simple microstability analysis. However, for 
reference, we note that the pre-pellet mixing length estimate of the diffusivity from TEMs (which cause particle transport) at the reference surface $\sqrt{\psi_N}=0.69$ 
gives $D\sim 1-2$ m$^2$s$^{-1}$. A similar estimate immediately after pellet injection ($t=t_0+t_{abl}$) at the surface $\sqrt{\psi_N}=0.85$
gives $D\sim 2$ m$^2$s$^{-1}$. These values are similar to  estimates of the effective diffusivity based on interpretative transport analysis,
which are in the order of 1.5 m$^2$s$^{-1}$ \cite{Valovic13}. It is worth noting that calculations of the neoclassical ion particle diffusivity in the region interested 
by the pellet perturbation are in the order of $D_{nc}\sim 1.5-2 \cdot 10^{-2}$ m$^2$s$^{-1}$ \footnote {On the other hand, the ion heat transport in spherical tokamaks 
has often been reported to be consistent with the neoclassical value \cite{Connor95,Roach01,Akers03}. This is also true for the plasmas analysed in this paper.}.

Challenging nonlinear simulations will be required to determine the transport coefficients more reliably and these must use 
domains that are sufficiently broad to capture the radial correlation length of the turbulence. For ion scale turbulence, this might typically extend to the 
order of 10 $\rho_i$, which approaches the radial size of the pellet induced density perturbation ($\rho_i=1.16$ cm on the surface at $\sqrt{\psi_N}=0.69$). 
This is likely to require global simulations, as the equilibrium gradient length scales will vary considerably over such a wide domain. 

Nevertheless, figures~\ref{GS2res1} and \ref{GS2res3}, respectively, demonstrate that TEMs are fully stable on the inner surface and 
linearly unstable on the outer surface after the pellet deposition. TEMs cause particle transport and this asymmetry in TEM stability 
around the peak is likely to enhance outward particle transport. This supports one-dimensional transport simulations performed for other tokamaks 
(see for example \cite{Garzotti03}), which indicate that the relaxation of the pellet deposition profile takes place on time scales faster than 
would be expected on the basis of unperturbed transport coefficients and points towards increased particle losses caused by the pellet perturbation.

\begin{figure}[htbp]
\begin{center}
\includegraphics[width=12cm]{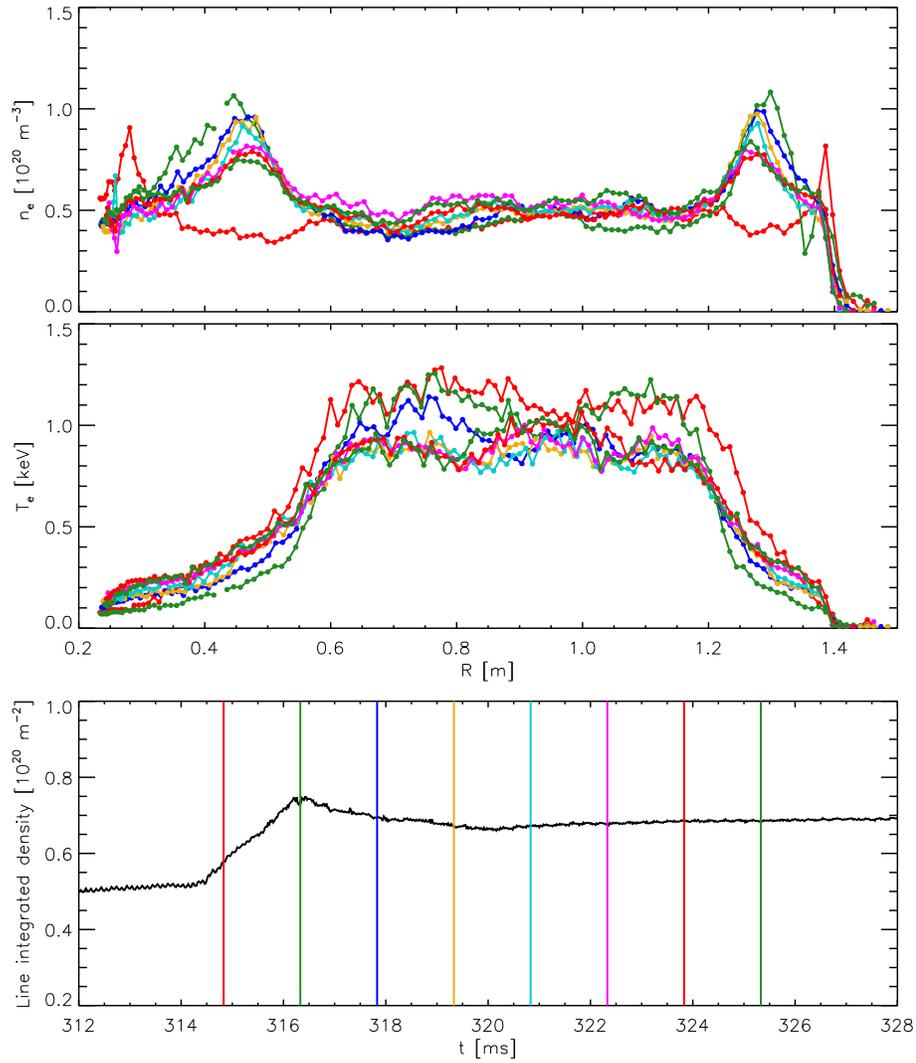}
\caption{\label{timing}Evolution of the electron density and temperature profiles and of the line integrated density 
during pellet injection for MAST shot 24743. The vertical bars indicate the times of the Thomson scattering density and temperature 
profile measurements. The profiles at 317.8 ms and 323.8 ms were chosen for the post-pellet GS2 microstability analysis.}
\end{center}
\end{figure}

\begin{figure}[htbp]
\begin{center}
\includegraphics[width=12cm]{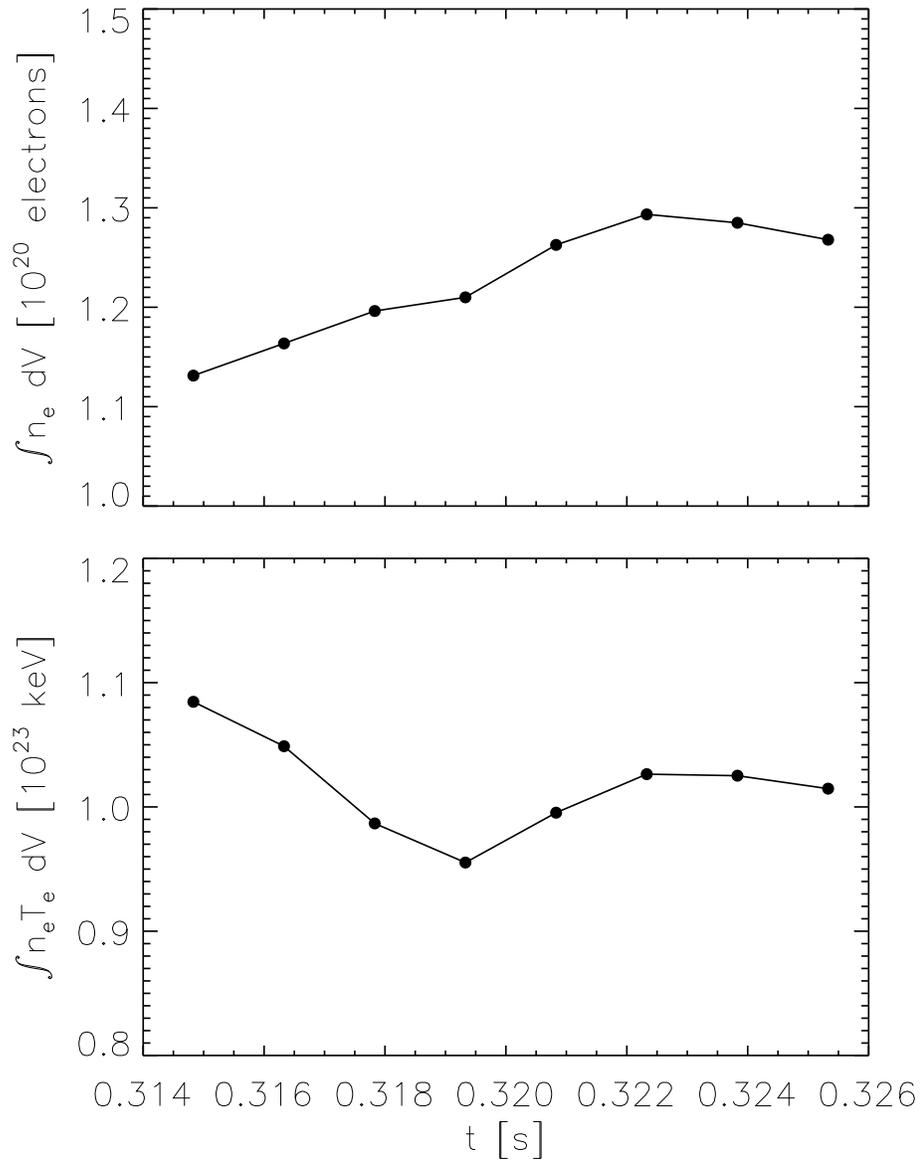}
\caption{\label{neint}Time evolution of the total number of electrons (top) and electron thermal energy (bottom) inside the flux 
surface corresponding to $\sqrt{\psi_N}=0.69$ for MAST shot 24743.}
\end{center}
\end{figure}

The dynamics described above can be seen in the evolution of the density profiles illustrated in figures \ref{timing} and \ref{exthomson} where it can be 
observed that, corresponding to the relaxation of the pellet deposition profile and the global decrease of the average plasma density, there is little increase of the 
central density and the pellet particles diffuse preferentially towards the plasma edge and are lost across the plasma separatrix. This is illustrated, 
in a more quantitative way, in figure \ref{neint}, showing the time evolution of the total number of electrons and electron thermal energy inside the flux surface 
corresponding to $\sqrt{\psi_N}=0.69$. It can be seen that, after the end of the pellet ablation (corresponding to the second time point), the total number of electrons  
inside $\sqrt{\psi_N} = 0.69$ increases by $1.45 \cdot 10^{19}$ between $t=315.0$ ms and $t=322.0$ ms. The numbers of particles injected by the beam and by the pellet during 
this period were $2.2 \cdot 10^{18}$ and $1.9 \cdot 10^{20}$ respectively. Therefore, the increase in the number of electrons inside $\sqrt{\psi_N} = 0.69$ can
be accounted for by only 6.5\% of the particles from the pellet diffusing into the core, indicating that the pellet particles diffuse in an asymmetric way, preferentially 
towards the edge of the plasma.
In this particular case, however, inspection of the evolution of the density profiles in the outer zone, where $dn/dr <0$, shows that the density decreases 
mainly during the time interval between 316.0 ms and 321.0 ms during which a burst of ELMs is observed. This is also seen on the line integrated density in figure 
\ref{timing}. Particle transport in the outer zone is influenced by both ELMs and micro-instabilities.

On the other hand, figure \ref{neint} shows weak evolution in the thermal electron stored energy within $\sqrt{\psi_N}=0.69$, which suggests that 
the electron heat flux is not strongly suppressed at this surface after the pellet ablation at $t=317.8$ ms. This appears to be inconsistent with the full suppression 
of local microinstabilities suggested by figure \ref{GS2res1}. It is conceivable that our local gyrokinetic analysis lacks important global effects, and we will 
extend our analysis in the future to try to resolve this question.

Our analysis also suggests that the shape of the pellet deposition profiles could play an important role in the evolution of the post-pellet density 
profile. Indeed, if the pellet ablation dynamics is such as to produce steep density gradients at the propagation front of the pellet density perturbation,
turbulence could become suppressed and the penetration of the pellet material slowed down because of the reduction in turbulence induced transport. On the other hand, 
this effect could be less important if, due to fast transport mechanisms affecting the motion of the pellet cloud \cite{Rozhansky04}, the pellet material is 
distributed on longer spatial scales and steep density gradients are avoided.

This last point shows the importance of interpreting the results from present day experiments in the light of physics-based models for particle transport
and analysing the microstability properties of the post-pellet density profiles when simulating pellet based fuelling schemes for future devices. 
Physics-based transport models still need validation and verification in order to be applied predictively to transient phenomena 
like pellet injection and, unfortunately, few experiments are equipped with diagnostics suitable to measure the evolution of the post-pellet profiles with the degree 
of accuracy needed for detailed microstability analysis.

Finally, to stress even further the relevance of the analysis presented in this paper, it is worth noting that in future generation devices, or even in 
a reactor plasma, where ELMs will have to be stabilized or at least considerably mitigated and MHD avoided, microturbulence could become the dominant player 
in determining the redistribution of the fuel injected into the plasma. First experiments on simultaneous pellet fuelling and ELM mitigation from MAST were 
reported in \cite{Valovic13}.

\section{Conclusions}\label{conclusions}
In this paper we have described the microstability analysis of a typical MAST pellet fuelled discharge. The accurate measurement 
of the post-pellet density and temperature profiles, made possible by the high spatial and temporal resolution of the MAST Thomson
scattering system, has allowed us to analyze three plasma regions around the pellet deposition peak. The results show that the microstability 
properties of the post-pellet profiles are highly sensitive to the rapid and large excursions in the gradients, $\beta_e$ and 
collisionality induced by the pellet injection.

In particular, at the innermost location, corresponding to the part of the pellet deposition profile where the density gradient is positive, all modes 
are stabilised by the reduction in $\eta_i$ and $\eta_e$ (induced by the increased density gradient) and by the favourable magnetic drifts. 
The picture for the pellet ablation peak (where the density gradient is small or negative) and beyond this radius is more complex. For low $k_y \rho_i$, ITG modes, 
which dominate before pellet injection, are stabilized, either by increased collisionality, at the pellet ablation peak where $dn/dr \approx 0$, or by reduced $\eta_i$, 
where $dn/dr < 0$. At the same time MTMs, which are unstable but subdominant before pellet injection, become dominant and can be driven more unstable by the 
increased $\beta_e$. After pellet injection, TEMs are driven unstable by the increased density gradient at the outer surface, where drifts are unfavourable, 
(whereas TEMs were stabilised on the inner surface where the drift opposes the electron diamagnetic flow velocity). Finally, ETG modes remain unstable at the peak of the 
ablation profile but are stabilised at the outer surface where $dn/dr < 0$ by the reduced $\eta_e$.

From the fuelling point of view this would imply that, if the density gradients inside the ablation peak are too steep, an asymmetric
evolution of the density profile should be expected whereby the anomalous diffusion of the pellet material toward the plasma core would be 
suppressed by the stabilization of the microturbulence and the pellet particles would diffuse preferentially toward the plasma edge.

The results of this study show that it is important to diagnose carefully the evolution of the density and temperature profiles 
in present day experiments in order to unveil the role of microinstabilities and to assess the performance of physics-based transport models.
This progress will be essential to build confidence in our modelling based extrapolations of pellet fuelled scenarios in future experiments
and to optimise the design of effective pellet fuelling.

\ack
The authors would like to thank Dr. J. W. Connor, Dr. D. Dunai, Dr. A. Field and Dr. Y.-c. Ghim for useful discussions and
are grateful for access to the supercomputer HECToR (EPSRC grant EP/H002081/1) used to perform the simulations.
This work was funded by the RCUK Energy Programme under grant EP/I501045 and the European Communities under the contract of 
Association between EURATOM and CCFE. The views and opinions expressed herein do not necessarily reflect those of the European 
Commission.

\newpage

\end{document}